\documentclass[12pt]{iopart}
\usepackage[utf8]{inputenc}
\begin{document}

\title{New Role of Null Lagrangians in Derivation of Equations of Motion
for Dynamical Systems}

\author{R. Das$^1$ and Z.E. Musielak$^2$} 

\address{$^1$Department of Physics and Earth Space Science, The University 
of Indianapolis, Indianapolis, IN 46227, USA}
\address{$^2$Department of Physics, The University of Texas at Arlington, 
Arlington, TX 76019, USA}

\ead{dasr@uindy.edu; zmusielak@uta.edu}

\begin{abstract}
The space of Null Lagrangians is the least investigated territory in dynamics since they are identically sent to zero by their Euler-Lagrange operator and thereby having no effects on equations of motion. A humble effort to discover the relevance of these Null Lagrangians in dynamics is made by introducing a generalized procedure (with respect to the recent procedure introduced by the authors of this paper) that takes advantage of the null-ness of these Lagrangians to construct non-standard Lagrangians that represent a range of interesting dynamical systems. 
By using the generalized procedure, derivation of equations of motion for a 
harmonic oscillator as well as for the Bateman and Duffing oscillators is 
presented. The obtained results demonstrate a new role played by the null 
Lagrangians and their corresponding non-standard Lagrangians in describing 
linear and nonlinear, and dissipative and non-dissipative dynamical systems. 
\end{abstract}

\maketitle

\section{Introduction} 

Dynamical systems with holonomic constraints can be analyzed using the Lagrangian
formalism. The foundation of this formalism is the smooth configuration manifold
$\mathcal{Q}$ constructed from the generalized coordinates of the system of interest 
with holonomic constraints. An additional structure, a tangent bundle $T\mathcal{Q}$, 
on $\mathcal{Q}$ is necessary to define Lagrangians for dynamical systems.  Then a 
Lagrangian $\mathcal{L}$ for a system is commonly defined as a $C^{\infty}$ map 
from $T\mathcal{Q}$ to the real line $\mathbf{R}$, i.e., $\mathcal{L}:T\mathcal{Q}
\rightarrow \mathbf{R}$ [1, 41]. 

In the case of one-dimensional dynamical systems, a Lagrangian can be symbolized 
as $\mathcal{L} = \mathcal{L}(x,\dot{x})$ in an open neighborhood $\mathcal{U}
\subset T\mathcal{Q}$ containing a point $p\in T\mathcal{Q}$ in a local coordinate 
system defined by the coordinate function $x:\mathcal{Q}\rightarrow \mathbf{R}$ 
and its associated tangent vector $\dot{x}$ at the point $p$  using a diffeomorphism 
$\phi :p\in \mathcal{U}\subset{T{\mathcal{Q}}} \rightarrow \mathcal{Q} \times 
\mathbf{R}$ known as a local trivialization.  Similarly, the explicit dependence of 
a Lagrangian on time can be defined as $\mathcal{L}:T\mathcal{Q}\times \mathbf
{R}\rightarrow \mathbf{R}$ and thus expressing $\mathcal{L} = \mathcal{L}(x,
\dot{x},t)$. After a suitable Lagrangian for an interested system is constructed or 
discovered, the time evolution of the system is analyzed by applying the Euler-Lagrange 
operator on the chosen Lagrangian, i.e., $\hat {EL} [ \mathcal{L} (\dot x, x, t) ] = 0$, 
with $\hat {EL}$ being the Euler-Lagrange (E-L) operator [1-4, 41].

There are two families of $\mathcal{L} (\dot x, x, t)$ that are used to obtained equations 
of motion, namely, standard and non-standard Lagrangians. The main characteristic of 
standard Lagrangians is the presence of the kinetic and potential energy-like terms [5], 
and these Lagrangians are commonly used in Classical Mechanics (CM) to derive equations 
of motion for different dynamical systems (e.g., [1-5]).  On the other hand, in non-standard 
Lagrangians, neither the kinetic nor the potential energy-like terms can be identified [6], 
nevertheless, they may give the same equations of motions of motion as the standard 
Lagrangians [7-10].    

The third family of Lagrangians is significantly different from the two families as it 
contains null Lagrangians (NLs), which identically satisfy the E-L equation and can 
be expressed as the total derivative of any scalar function, called here a gauge function.  
The NLs have been extensively studied in mathematics (e.g., 11-18]); however, since 
they do not give equations of motion, their physical applications have been least explored and  
limited to only a few cases that involed elasticity [19,20] and forces in CM [21,22].     

An interesting relationship was recently found between the non-standard and null 
Lagrangians [23] and it was shown that 
\begin{equation}
 \frac{d}{dt}\left[L_{null} (\dot x, x, t)\right] =0,
\label{eom-null}
\end{equation}
gives an equation of motion for any null Lagrangian $L_{null} (\dot x, x, t)$.  In 
other words, the condtion plays the same role for NLs as the E-L equation plays 
for standard and non-standard Lagrangians.  

The main advantage of Eq. (\ref{eom-null}) is that its form is much simpler than 
the E-L equation and thus it is easier to derive an equation of motion.  However, 
its disadvantage is that the resulting equation of motion must be in a form that 
obeys a special relationship between the coefficients of this equation; the 
relationship is unique for a given dynamical system but varies for different 
systems [23].  Therefore, the mainpurpose of this paper is to identify limitations 
of the method and Eq. (\ref{eom-null}), and find a way to remove these limitations 
and develop a generalized procedure that allows deriving equations of motion that 
are unrestricted by any relationship between its coefficients.  Our paper presents 
this generalized procedure and applies it to specific examples of dynamical systems 
such as a harmonic oscillator as well as the Bateman [24] and Duffing [25,26] 
oscillators, which shows a broad range of applications of the developed procedure. 

The paper is organized as follows: Section 2 gives an overview of previously 
developed method and its limitation; in Section 3, construction of non-standard 
Lagrangians from null Lagrangians is described; Section 4 presents applications
of the obtained results to different osillatory systems, including a harmonic 
oscillator and the Bateman and Duffing oscillators; our conclusions are given 
in Section 5. 

\section{Previously developed method and its limitations}
\label{formalism}

Null Lagrangians, $L_{null} (\dot x, x, t)$, are singular Lagrangians that satisfy
$\hat {EL} [ L_{null} (\dot x, x, t) ] = 0$.  In other words, all null Lagrangians are 
identically sent to zero by the Euler-Lagrange ($\hat EL$) operator.  The obvious 
ramification is that no unique trajectory of the time evolution of the corresponding 
system is derivable through the extremization of the action functional defined in 
terms of the NLs.  However, it is possible to exploit these NLs for constructing 
non-standard Lagrangians to represent certain interesting dynamical systems.  
Recently, a procedure for deriving equations of motion for a given NL has been 
developed [23], and in the following we briefly summarize the procedure by 
presenting its main condition.  Then, we show its limitations and a need for 
its generalization.

Let $L_{null}(\dot{x},x,t): \mathcal{U} \subset{T\mathcal{Q}}\rightarrow 
\mathbf{R}$ be an invertible and $C^{\infty}$ function.  Then a non-standard 
Lagrangian $\mathcal{L}:  \mathcal{U} \subset{T\mathcal{Q}} \rightarrow \mathbf{R}$ 
can be defined by using at least a $C^2$ function $F: \mathbf{R} \rightarrow \mathbf{R}$ 
as $\mathcal{L}(\dot{x},x,t)=(F \circ L_{null})(\dot{x},x,t))=F( L_{null}(\dot{x},x,t))$ 
such that, for any permissible form of function $\mathcal{L}(\dot{x},x,t)= (F\circ L_{null})
(\dot{x},x,t)$ satisfying $p_{{null}} \left[\frac{d^{2} F}{d L_{null}^{2}} \right]\ne 0$, 
the corresponding equation of motion for a dynamical system represented by $\mathcal{L}
(\dot{x},x,t)$ is given by Eq. (\ref{eom-null}) with $p_{null}:= \partial {L}_{null}/\partial 
\dot{x}$.

The above construction can be carried out by using a null Lagrangian, ${L}_{null}(\dot{x},x,t)
=d\Phi(x,t)/dt$, constructed from a smooth and invertible gauge function $\Phi(x,t)$ defined 
over ${\Omega}\in \mathcal{Q}\times \mathbf{R}$.  It is straightforward to show that Eq. 
(\ref{eom-null}), upon inserting $L_{null}(\dot{x},x,t)=d\Phi(x,t)/dt$, takes the form:
\begin{equation}
\Phi^{'}(x,t) \ddot{x} +\left[\Phi^{''}(x,t)\dot{x}+2(\dot{\Phi})^{'}(x,t)\right]\dot{x}+
\ddot{\Phi}(x,t) =0,
\label{phinulleom}
\end{equation}
where primes and dots are defined respectively as $\Phi^{'}:=\partial \Phi(x,t)/\partial x$, 
$\Phi^{''}:=\partial^{2} \Phi(x,t)/\partial x^{2}$, $\dot{\Phi} :=\partial \Phi(x,t)/\partial t$ 
and $\ddot{\Phi} :=\partial^{2} \Phi(x,t)/\partial t^{2}$.\\

To apply this result to dynamical systems, let us consider the following general equation 
of motion 
\begin{equation}
\ddot{x} + \alpha (x,t) \dot{x}^2 + \beta (x,t) \dot{x} + \gamma (x,t) x = 0,
\label{generaleom}
\end{equation}
where $\alpha (x,t)$, $\beta (x,t)$ and $\gamma (x,t)$ are the coefficients determined 
by a given dynamical system.  For specific forms of these coefficients, this equation of 
motion describes a broad variety of dynamical systems ranging from a harmonic oscillator 
and the Bateman and Duffing oscillators to other dynamical systems, including the population
dynamics models, such as the Lotka-Volterra, Verhulst, Gompertz and Host-Parasite models,
and others [27]. 

By comparing Eq. (\ref{phinulleom}) to Eq. (\ref{generaleom}), we find 
\begin{equation}
\Phi^{''}(x,t) = \alpha (x,t) \Phi^{'}(x,t), 
\label{phi1}
\end{equation}
\begin{equation}
\dot{\Phi}^{'}(x,t) = \frac{1}{2} \beta (x,t) \Phi^{'}(x,t), 
\label{phi2}
\end{equation}
and 
\begin{equation}
\ddot{\Phi} (x,t) = \gamma (x,t) x \Phi^{'}(x,t), 
\label{phi3}
\end{equation}
with $\Phi^{'}(x,t) \neq 0$.  These are the conditions that the gauge function 
must satisfy in order to give the null Lagrangian for Eq. (\ref{generaleom}).  
Clearly, the fact that different derivaties of $\Phi (x,t)$ depend on the coefficients 
$\alpha (x,t)$, $\beta (x,t)$ and $\gamma (x,t)$ implies that the resulting gauge 
function, which is the solution for these conditions, must depend on a certain 
relationship between the coefficients.  In other words, $\Phi (x,t)$ that satisfy
these conditions and the NL resulting from this gauge function will give only
the equation of motion for which the relationship is valid.    

After eliminating $\Phi (x,t)$ from the above conditions, we obtain the following
relationship between the coefficients
\begin{equation}
\frac{1}{2} \left [ \dot {\beta} (x,t) + \frac{1}{2} \beta^2 (x,t) \right ] = 
\gamma^{'} (x,t) x + \gamma (x,t) \left [ 1 + \alpha (x,t) x \right ]. 
\label{relation-gen}
\end{equation}
If this condition is satisfied for a given equation of motion, then the gauge 
function and the corresponding null Lagrangian can be  constructed, and 
when the latter is substituted into Eq. (\ref{eom-null}), the original 
equation of motion is obtained.   The condition significantly restricts 
forms of the equations of motion that can be derived using this method.
In the following, we first consider some special restricted cases and then
identify unrestricted cases.

In case $\alpha (x,t) = \alpha (x)$, $\beta (x,t) = \beta (x)$ and 
$\gamma (x,t) = \gamma(x)$, the condition reduces to 
\begin{equation}
\frac{1}{4} \beta^2 (x) = \gamma^{'} (x) x + \left [ 1 + \alpha (x) x \right ] 
\gamma (x),
\label{relation1}
\end{equation}
which shows the dependence between all the coefficients.

If $\alpha (x,t) = \alpha (t)$, $\beta (x,t) = \beta (t)$ and $\gamma (x,t) 
= \gamma(t)$, the condition becomes 
\begin{equation}
\frac{1}{2} \left [ \dot {\beta} (t) + \frac{1}{2} \beta^2 (t) \right ] = 
\gamma (t),  
\label{relation2}
\end{equation}
which demonstrates the relationship between $\beta (t)$ and $\gamma (t)$.
In this case, $\alpha (t ) = 0$ is required to eliminate the $x$-dependence
from the condition.  

Now, the case of constant coefficients, $\alpha (x,t) = \alpha_o$ = const, 
$\beta (x,t) = \beta_o$ = const and $\gamma (x,t) = \gamma_o$ = const,
requires also that $\alpha_o = 0$, for the same reason as above, and gives
the relationship between $\beta_o$ and $\gamma_o$ to be $\beta_o^2 = 
4 \gamma_o$.

Another interesting case is when $\alpha (x,t) = \gamma (x,t) = 0$ and 
$\beta (x,t) = \beta (t)$.  Then, we find
\begin{equation}
\dot {\beta} (t) + \frac{1}{2} \beta^2 (t) = 0,  
\label{relation3}
\end{equation}
with the solution
\begin{equation}
\beta (t) = \frac{2}{( c_1 + t)^2},
\label{sol-beta}
\end{equation}
where $c_1$ is an integration constant. The obtained result means that 
the gauge function and null Lagrangian exist only for the equation of 
motion with $\beta$ coefficient having this time dependence; for some 
other special cases see [23].

Let us now identify more interesting cases that are not restricted by the 
condition given by Eq. (\ref{relation-gen}).  Obviously, the simplest such 
case is when $\alpha (x,t) = \beta (x,t) = \gamma (x,t) = 0$, which 
automatically satisfies the condition.  Then, the gauge function that is 
the solution to Eqs. (\ref{phi1}),  (\ref{phi2}) and (\ref{phi3}) becomes
\begin{equation}
\Phi (x,t) = c_1 x + c_2 t + c_3,
\label{phi_inertia}
\end{equation}
where $c_1$, $c_2$ and $c_3$ are integration constants, and the 
corresponding null Lagrangian is 
\begin{equation}
L_{null} (x) = c_1 \dot x + c_2,
\label{null_inertia}
\end{equation}
and its substitution into Eq. (\ref{eom-null}) gives the {\it law of inertia} 
$\ddot x = 0$.   

By taking $\alpha (x,t) = \beta (x,t) = 0$ and $\gamma (x,t) \neq 0$,
the condition reduces to
\begin{equation}
\gamma^{'} (x,t) x + \gamma (x,t) = 0,  
\label{relation-gamma}
\end{equation}
with the solution
\begin{equation}
\gamma (x,t) = \frac{f_0 (t)}{x},  
\label{solution-gamma}
\end{equation}
where $f_0(t)$ is an arbitrary function.  Then, the gauge function is
\begin{equation}
\Phi  (x,t) = c_1 x + c_1 \int^t \left [ \int^{\bar t} f_0 (\tilde t ) d \tilde {t}
\right ] d \bar t,  
\label{phi-second}
\end{equation}
and the null Lagrangian is given by
\begin{equation}
L_{null} (x,t) = c_1 \dot x + c_1 \int^{t} f_0 (\tilde t ) d \tilde {t}.  
\label{null-second}
\end{equation}
Substitution of this Lagrangian into Eq. (\ref{eom-null}) gives the {\it second
law of dynamics} $\ddot x = F (t)$, where $F(t) = - f_0(t)$. 

The last unrestricted case is when $\alpha (x,t) \neq 0$ and $\beta (x,t) = 
\gamma (x,t) = 0$.  Then, the gauge function is
\begin{equation}
\Phi (x) = c_1 \int^{x} \e^{I_{\alpha} (\tilde x)} d \tilde x + c_2 t + c_3\ ,
\label{phi-damp}
\end{equation}
and the null Lagrangian becomes 
\begin{equation}
L_{null} (\dot x,x) = c_1 \dot x \e^{I_{\alpha} (x)} + c_2\ ,
\label{null-damp}
\end{equation}
where $I_{\alpha} (x) = \int^x \alpha (\bar x) d \bar x$.

By substituting $L_{null} (\dot x,x)$ into Eq. (\ref{eom-null}),
we obtain
\begin{equation}
\ddot{x} + \alpha_2 (x) \dot{x}^2 = 0\ ,
\label{eom-damp}
\end{equation}
which is the equation of motion for an oscillatory system with the quadratic 
damping [23, 9].

The above results demonstrate that the previously developed method can 
be used to derive the first and second law of Newtonian dynamics by using 
the constructed gauge functions and their corresponding null Lagrangians.  
Moreover, the method allows obtaining the equation of motion for an 
oscillatory system with the quadratic damping term.  The derivations of 
these three equations are unrestricted by any constraints.  However, any 
application of the previous method to other dynamical systems including 
harmonic and linearly damped oscillators is restricted by special relationships 
between the system's coefficients.  Therefore, in order to apply to a broad variety of dynamical systems, a generalization of the previous method is needed and this is the main purpose of this paper.  As shown in the following, the generalized method allows constructing gauge functions 
and null Lagrangians and deriving equations of motion for a harmonic 
oscillator as well as for the Bateman and Duffing oscillators. 

\section{Construction of non-standard Lagrangians from null Lagrangians}

After the above construction of the non-standard Lagrangians from null Lagrangians, the following generalization is attempted to obtain new non-standard Lagrangians to represent more dynamical systems of interest. 

{\bf Proposition 1:} Let $F(L_{null}(\dot{x},x,t))$, $G(L_{null}(\dot{x},x,t))$ and $M(L_{null}(\dot{x},x,t))$  : $\mathcal{U} \subset{T\mathcal{Q}} \times \mathbf{R}  \rightarrow \mathbf R$ be at least $C^{2}$ functions of a certain invertible $L_{null}\in \mathcal{F}$ along with invertible and smooth maps $Q(\dot{x},x,t)$, $R(\dot{x},x,t)$  and $U(\dot{x},x,t):$  $\mathcal{U} \subset{T\mathcal{Q}} \times \mathbf{R}\rightarrow \mathbf R$ satisfying constraints $Q(\dot{x},x,t)=R(\dot{x},x,t) L_{null}(\dot{x},x,t)$ and $Q \frac{d F}{d L_{null}}+R \frac{d G}{d L_{null}}= \lambda (constant)$. Then there exists a non-standard Lagrangian of the form 
\[
\mathcal{L}_{ns}(\dot{x},x,t)=  Q(\dot{x},x,t)F(L_{null}(\dot{x},x,t))
\]
\begin{equation}
+R(\dot{x},x,t)G(L_{null}(\dot{x},x,t))+U(\dot{x},x,t)M(L_{null}(\dot{x},x,t))
\label{non-standard-lagrangian}
\end{equation}
such that
the equation of motion of this non-standard Lagrangian is given by
\[
\hat{EL}[RL_{null}(\dot{x},x,t)]F(L_{null}) +  \hat{EL}[R(\dot{x},x,t)]G(L_{null}) +\hat{EL}[U(\dot{x},x,t)]M(L_{null})
\]
\[
+
\left[\lambda \left[\frac{p_{R}}{R}+\frac{p_{null}}{L_{null}}\right]+p_{U}\frac{dM}{dL_{null}}\right] \frac{dL_{null}}{dt}
\]
\begin{equation}
+p_{null}\frac{d}{dt}\left[U\frac{dM}{dL_{null}}\right]-p_{{null}}\frac{R}{L_{null}}\frac{dG}{dt}= 0,
\label{extended-eom-for-general-lagrangian}
\end{equation}
with $p_{U}:=\partial U/\partial \dot{x}$, $p_{null}=\partial L_{null}/\partial \dot{x}$ and $p_{R}=\partial R/\partial \dot{x}$.

{\bf Proof:} The derivation of Eq. (\ref{extended-eom-for-general-lagrangian}) follows directly from $\hat{EL}[\mathcal {L}_{ns}]=0$, and the constraints $Q(\dot{x},x,t)=R(\dot{x},x,t) L_{null}(\dot{x},x,t)$ and $Q \frac{d F}{d L_{null}}+R \frac{d G}{d L_{null}}= \lambda (constant)$. More precisely, the action of the Euler-Lagrange operator on the non-standard Lagrangian, Eq. (\ref{non-standard-lagrangian}), yields
 \[
 \left[p_{Q}\frac{dF}{dL_{null}}+p_{R}\frac{dG}{dL_{null}}+p_{U}\frac{dM}{dL_{null}}\right]\frac{dL_{null}}{dt}
 \]
 \[
 +p_{{null}}\frac{d}{dt}\left[Q \frac{d F}{d L_{null}}+R \frac{d G}{d L_{null}}+U \frac{d M}{d L_{null}}\right]
 \]
\begin{equation}
 + \hat{EL}[Q(\dot{x},x,t)]F(L_{null}) +  \hat{EL}[R(\dot{x},x,t)]G(L_{null})+  \hat{EL}[U(\dot{x},x,t)]M(L_{null})= 0.
\label{eom-for-general-lagrangian}
\end{equation}
Then the application of the above constraints reduces the above equation, Eq.(\ref{eom-for-general-lagrangian}), to Eq. (\ref{extended-eom-for-general-lagrangian}). This completes the proof.

Here, a few noteworthy remarks are in order. First, it follows directly from the constraint $Q \frac{d F}{d L_{null}}+R \frac{d G}{d L_{null}}= \lambda$ that
\begin{equation}
 F(L_{null})= \int{\frac{\lambda }{RL_{null}}dL_{null}}-\int{\frac{1 }{L_{null}}\left[\frac{d G}{d L_{null}}\right] dL_{null}} + constant,
\label{FG-relation}
\end{equation}
which implies that the knowledge of the form of $G$ and the constant $\lambda$ completely determines $F$ modulo a constant provided Eq. (\ref{FG-relation}) is integrable. Second, the constraint $Q(\dot{x},x,t)=R(\dot{x},x,t) L_{null}(\dot{x},x,t)$ ensures that the above construction of non-standard Lagrangians centers around only null Lagrangians if the last term in Eq. (\ref{non-standard-lagrangian}) vanishes. In this case, the equation of motion, Eq. (\ref{extended-eom-for-general-lagrangian}) simplifies, for $\lambda=0$ to 
\begin{equation}
 \hat{EL}[RL_{null}(\dot{x},x,t)]F(L_{null}) +  \hat{EL}[R(\dot{x},x,t)]G(L_{null}) -p_{{null}}\frac{R}{L_{null}}\frac{dG}{dt}= 0.
\label{simplified-eom-for-general-lagrangian2}
\end{equation}
The above simplified equation of motion, Eq. (\ref{simplified-eom-for-general-lagrangian2}), turns out to be extremely advantageous for deriving the non-standard Lagrangians for several well-known dynamical systems, which will be presented in the next section. Last, even for the case of a non-vanishing last term with $M(L_{null})=m(constant)$, the above construction in Proposition 1 simplifies  to the following corollary.

{\bf Corollary 1:} Let the construction in Eq. (\ref{non-standard-lagrangian}) be simplified as follows
\[
\mathcal{L}_{ns1}(\dot{x},x,t)=  Q(\dot{x},x,t)F(L_{null}(\dot{x},x,t))
\]
\begin{equation}
+R(\dot{x},x,t)G(L_{null}(\dot{x},x,t))+mU(\dot{x},x,t),
\label{FGM-lagrangian}
\end{equation}
where $m$ is a constant and $Q$, $R$, $U$, $F(L_{null})$ and $G(l_{null})$ are defined as above in Proposition 1 satisfying the constraints  $Q(\dot{x},x,t)=R(\dot{x},x,t) L_{null}(\dot{x},x,t)$ and $Q \frac{d F}{d L_{null}}+R \frac{d G}{d L_{null}}= \lambda (constant)$. Then it follows that the  corresponding equation of motion for $\mathcal{L}_{ns1}(\dot{x},x,t)$ turns out to be 
\[
\hat{EL}[RL_{null}(\dot{x},x,t)]F(L_{null}) +  \hat{EL}[R(\dot{x},x,t)]G(L_{null}) +m\hat{EL}[U(\dot{x},x,t)]
\]
\begin{equation}
+
\lambda \left[\frac{p_{R}}{R}+\frac{p_{null}}{L_{null}}\right]\frac{dL_{null}}{dt}
-p_{{null}}\frac{R}{L_{null}}\frac{dG}{dt}= 0,
\label{potential-eom-for-general-lagrangian}
\end{equation}

The proof follows directly from Eq. (\ref{extended-eom-for-general-lagrangian}). Furthermore, if $U(\dot{x},x,t)=U(x,t)$ is a function of position and time only, then $\hat{EL}[U(\dot{x},x,t)]=-\partial U/\partial x$ behaves like a force term, which will be of particular interest in our application in the following section. Next, it is crucial to explore the scope of this novel construction of non-standard Lagrangians from null-Lagrangians to represent interesting dynamical systems.

\section{Applications to dynamical systems}

\subsection{General approach}

The application of the above construction in Proposition 1 to derive non-standard Lagrangians for dynamical systems becomes convenient if a null-Lagrangian is expressed as follows
\begin{equation}
L_{null}=\frac{d\Phi (x,t)}{dt}=\Phi^{'}\dot{x}+\dot{\Phi}=\left(\dot{x}+\frac{\dot{\Phi}}{\Phi^{'}}\right)\Phi^{'},
\end{equation}
where $\Phi^{'}:=\partial \Phi/\partial x \ne 0$ and $\dot{\Phi}:=\partial \Phi/\partial t$. Now, $Q= RL_{null}$ implies that $Q=\left(\dot{x}+\frac{\dot{\Phi}}{\Phi^{'}}\right)$ and $R=1/\Phi^{'}$. Then, with the vanishing last term in Proposition 1,  the equation of motion, Eq. (\ref{simplified-eom-for-general-lagrangian2}), takes the following form 
\begin{equation}
\frac{\Phi^{''}}{(\Phi^{'})^{2}} G(L_{null})+ \left[\lambda \Phi^{'}-\frac{dG}{dL_{null}} \right]\frac{1}{L_{null}}\frac{dL_{null}}{dt}+\left[\frac{\dot{\Phi}\Phi^{''}-(\dot{\Phi})^{'}\Phi^{'}}{(\Phi^{'})^{2}}\right]F(L_{null})=0.
\label{eom}
\end{equation}
It is easy to notice that the last term in the above equation, Eq. (\ref{eom}), vanishes for a gauge function $\Phi$ independent of time explicitly, i.e., $\Phi(x,t)=\Phi(x)$.

At this juncture, the knowledge of the form of $G(L_{null})$ is necessary to move further.  The form of $G(L_{null})$ may be determined by casting the above equation of motion, Eq. (\ref{eom}), into the form of Eq. (\ref{generaleom}). Then a comparison with the equation of motion of a desired dynamical system such as an oscillatory system may be advantageous to determine the form of $G(L_{null})$. Hence, it follows from Eq. (\ref{phinulleom}) and $\lambda \Phi^{'}-\frac{dG}{dL_{null}} \ne 0$ that the equation of motion, Eq. (\ref{eom}), may be recast as
\[
 \ddot{x} + \frac{\Phi^{''}}{\Phi^{'}}\dot{x}^{2}+2\frac{(\dot{\Phi})^{'}}{\Phi^{'}}\dot{x}+
\frac{\Phi^{''}}{(\Phi^{'})^{3}} \left[\frac{L_{null}}{\lambda \Phi^{'}-\frac{dG}{dL_{null}}}\right]G(L_{null})
\]
\begin{equation}
+\left[\frac{\dot{\Phi}\Phi^{''}-(\dot{\Phi})^{'}\Phi^{'}}{(\Phi^{'})^{3}}\right]\left[\frac{L_{null}}{\lambda \Phi^{'}-\frac{dG}{dL_{null}}}\right]F(L_{null})+\frac{\ddot{\Phi}}{\Phi^{'}}=0.
\label{modifiedeom}
\end{equation}
Next, with the above modified equation of motion, Eq. (\ref{modifiedeom}), the construction of non-standard Lagrangians for oscillatory systems is explored below.

\subsection{Harmonic oscillator}

As the first desirable application to a harmonic oscillator, a comparison between Eq. (\ref{modifiedeom}) and the desirable equation of motion, $\ddot{x}+\gamma(x,t) f(x)=0$ with $\gamma (x,t)=\gamma_{0}$ (a constant), becomes easier if the choices $\Phi(x,t)=\Phi(x)$ and $\lambda = 0$ are made. Then Eq. (\ref{modifiedeom}) simplifies to 
\begin{equation}
 \ddot{x} + \frac{\Phi^{''}}{\Phi^{'}}\dot{x}^{2}-
\frac{\Phi^{''}}{(\Phi^{'})^{3}} \left[\frac{L_{null}}{\frac{dG}{dL_{null}}}\right]G(L_{null})=0,
\label{eomoscillator}
\end{equation}
and then the comparison between Eq. (\ref{eomoscillator}) and $\ddot{x}+\gamma_{0} f(x)=0$ yields
\begin{equation}
 \frac{\Phi^{''}}{\Phi^{'}}\dot{x}^{2}-
\frac{\Phi^{''}}{(\Phi^{'})^{3}} \left[\frac{L_{null}}{\frac{dG}{dL_{null}}}\right]G(L_{null})=\gamma_{0} f(x).
\label{comaprison}
\end{equation}
Next, it is obvious that
\begin{equation}
\frac{\Phi^{''}}{(\Phi^{'})^{3}} \left[\frac{L_{null}}{\frac{dG}{dL_{null}}}\right]G(L_{null})=\pm \frac{\Phi^{''}}{(\Phi^{'})^{3}} +  \frac{\Phi^{''}}{\Phi^{'}}\dot{x}^{2},
\label{assumption}
\end{equation}
will reduce Eq. (\ref{eomoscillator}) to $\ddot{x}\mp\frac{\Phi^{''}}{(\Phi^{'})^{3}}=0$, which in turn produces the constraint on the gauge function, $\Phi^{''}\pm\gamma_{0}f(x)(\Phi^{'})^{3}=0$ to yield the desirable equation of motion, 
$\ddot{x}+\gamma_{0} f(x)=0$. Now, from Eq. (\ref{assumption}), it easily follows that
\begin{equation}
\frac{dG}{dL_{null}}=\frac{L_{null}}{L^{2}_{null}\pm1}G,
\label{G-equation}
\end{equation}
where $L_{null}=\Phi^{'}\dot{x}$ is used. Next, for $+1$ and $-1$, it is straightforward to obtain the form of G from the above equation Eq. (\ref{G-equation}) as $G(L_{null})=\sqrt{1+L^{2}_{null}}+c_{0}$ and $G(L_{null})=\sqrt{L^{2}_{null}-1}+c_0$ respectively with $c_{0}$ being a constant of integration. However, since $G(L_{null})=\sqrt{L^{2}_{null}-1}+c_0$ is constrained to be valid for only $L_{null} > 1$, $G(L_{null})=\sqrt{1+L^{2}_{null}}+c_{0}$ is a natural choice for application and thus it is chosen in this paper.

After determining the form of $G(L_{null})$, it is necessary to determine the form of the appropriate null-Lagrangian $L_{null}$ to obtain the equation of motion for a dynamical system. For an oscillatory system, it follows from the above constraint, $\Phi^{''}+\gamma_{0}f(x)(\Phi^{'})^{3}=0$, that any integrable function $f(x)$ defined over a appropriate domain of the configuration space $\mathcal{Q}$ of the oscillator satisfying $\Phi^{'}=\left[c_{1}+2\gamma_{0}\int^{x}{f(\xi)d\xi}\right]^{-1/2}$, with $c_1$ being the constant of integration, will give the respective null-Lagrangian, $L_{null}=\Phi^{'}\dot{x}$.

As a simple check, it is easy to see that $f(x)=x$ after inserted in $L_{null}=\dot{x}\Phi^{'}=\dot{x} \cdot \left[c_{1}+2\gamma_{0}\int^{x}{f(\xi)d\xi}\right]^{-1/2}$ will generate the null-Lagrangian $L_{null}=\frac{\dot{x}}{\sqrt{\gamma_{0}x^{2}+c_1}}$ with $c_{1}$ being a constant of integration. Next, with the form of $\Phi^{'}= \frac{1}{\sqrt{\gamma_{0}x^{2}+c_1}}$,  the equation of motion, $\ddot{x}-\frac{\Phi^{''}}{(\Phi^{'})^{3}}=0$, reduces to the expected equation of motion for a simple harmonic oscillator, $\ddot{x}+\gamma_{0}x=0$. Moreover, the non-standard Lagrangian for a simple harmonic oscillator, from Proposition 1, is determined to be
\begin{equation}
L^{HO}_{ns}(\dot{x}, x, t)=\dot{x}\cdot {sinh}^{-1}\left(\frac{\dot{x}}{\sqrt{\gamma_{0}x^{2}+c_1}}\right)-\sqrt{\gamma_{0}x^{2}+c_1} \cdot \sqrt{1+\left(\frac{\dot{x}}{\sqrt{\gamma_{0}x^{2}+c_1}}\right)^2},
\label{HO}
\end{equation}
where $F(L_{null})=sinh^{-1}(L_{null}) + \  constant$ is used after obtaining from Eq. (\ref{FG-relation}). It should be noted here that $G(L_{null})=-\sqrt{1+L^{2}_{null}}+c_{0}$ is necessary to obtain $F(L_{null})=sinh^{-1}(L_{null}) + \  constant$ from Eq. (\ref{FG-relation}) to have Eq. (\ref{HO}). This is compatible with Proposition 1 since an overall extra negative sign appears naturally in the transition from Eq. (\ref{eom}) to Eq. (\ref{eomoscillator}). Finally, as a cross check, it can be easily shown that $\hat{EL}\left[L^{HO}_{ns}(\dot{x}, x, t)\right]=\ddot{x}+\gamma_{0} x=0$.

Similarly, it follows that after inserting $f(x)=sin(ax)$, with $a\in \bf R$, in $L_{null}=\dot{x}\Phi^{'}=\dot{x} \cdot \left[c_{1}+2\gamma_{0}\int^{x}{f(\xi)d\xi}\right]^{-1/2}$  generates the null-Lagrangian $L^{f}_{null}= \frac{\sqrt{a}\dot{x}}{\sqrt{2}\sqrt{ac_{2}-\gamma_{0}cos(ax)}}$ and the corresponding non-standard Lagrangian turns out to be
\begin{equation}
L^{f}_{ns}(\dot{x}, x, t)=\sqrt{a}\dot{x}\cdot {sinh}^{-1}(L^{f}_{null})- \sqrt{2}\sqrt{ac_{2}-\gamma_{0}cos(ax)}\cdot \sqrt{1+(L^{f}_{null})^2},
\label{HOf}
\end{equation}
with $c_2$ being a constant of integration and obviously the respective equation of motion is given by $\hat{EL}\left[L^{f}_{ns}(\dot{x}, x, t)\right]=\ddot{x}+\gamma_{0}sin(ax)=0$. It should be noted here that it is the form of the null Lagrangian by construction that determines the dynamical system of interest once the structure of a non-standard Lagrangian such as Eq. (\ref{HO}) and Eq. (\ref{HOf}) is chosen.

\subsection{Linear and non-linear Bateman oscillator}

Once a non-standard Lagrangian for an oscillator is obtained, it is possible to derive the related non-standard Lagrangian for a damped oscillator by multiplying the non-standard Lagrangian with a damping function $e^{bt}$ for a damping parameter $b$ as in [42,43]. However, in order to be consistent with the construction of non-standard Lagrangian above, i.e., Proposition 1, it is necessary to express the null-Lagrangian as
\begin{equation}
L_{null}=e^{bt}\left(\dot{x}+\frac{\dot{\Phi}}{\Phi^{'}}\right)\Phi^{'}e^{-bt}
\end{equation}
so that $L_{null}$ does not depend on time explicitly with transferring the explicit time dependence to $Q$ and $R$ as $Q=e^{bt}\left(\dot{x}+\frac{\dot{\Phi}}{\Phi^{'}}\right)$ and $R=e^{bt}/\Phi^{'}$.

With the above identifications along with $F=sinh^{-1}(L_{null})$  and $G=-\sqrt{1+L_{null}^2}$, a non-standard Lagrangian for a damped oscillator can be expressed as
\begin{equation}
L^{damped}_{ns}(\dot{x}, x, t)=e^{bt}\dot{x}\cdot {sinh}^{-1}(L^{f}_{null})- \frac{1}{\Phi^{'}e^{-bt}}\cdot \sqrt{1+(L^{f}_{null})^2}.
\label{damped-HOf}
\end{equation}
Then it is straightforward to obtain the corresponding equation of motion for a damped oscillator as
\begin{equation}
\ddot{x}+\frac{b}{\Phi^{'}} \cdot sinh^{-1}(L^{f}_{null}) \cdot \sqrt{1+(L^{f}_{null})^2}+\frac{\Phi^{''}}{\Phi^{'}}\dot{x}^{2}-\frac{\Phi^{''}}{(\Phi^{'})^3}\left(1+(L^{f}_{null})^2\right)=0.
\label{eom-damped-oscillator}
\end{equation}
However, in order to extract the desired equation of motion for a damped oscillator, it is necessary to express the second term in Eq. (\ref{eom-damped-oscillator}) using Taylor expansion around $L^{f}_{null}=0$ as
\begin{equation}
\frac{b}{\Phi^{'}} \cdot sinh^{-1}(L^{f}_{null}) \cdot \sqrt{1+(L^{f}_{null})^2} \approx b \dot{x} + \frac{1}{3}b(\Phi^{'})^{2}{\dot{x}^{3}}- \frac{2}{15} b(\Phi^{'})^{4}\dot{x}^{5}+\cdots,
\label{taylor-expansion}
\end{equation}
where $L^{f}_{null}=\phi^{'}\dot{x}$ is used. Inserting Eq. (\ref{taylor-expansion}) into Eq. (\ref{eom-damped-oscillator}) gives
\begin{equation}
\ddot{x}+\left( b \dot{x} + \frac{1}{3}b(\Phi^{'})^{2}{\dot{x}^{3}}- \frac{2}{15} b(\Phi^{'})^{4}\dot{x}^{5}+\cdots\right)-\frac{\Phi^{''}}{(\Phi^{'})^3}=0.
\label{expanded-eom-damped-oscillator}
\end{equation}

It is obvious that Eq. (\ref{expanded-eom-damped-oscillator}) reduces to the equation of motion for a damped oscillator in the zeroth order approximation, $L^{f}_{null} \ll 1$. In particular, with $-\frac{\Phi^{''}}{(\Phi^{'})^3}=\gamma_{0}x$, Eq. (\ref{expanded-eom-damped-oscillator}) takes the form of the linear Bateman oscillator, i.e., $\ddot{x}+b\dot{x}+\gamma_{0}x=0$ in the zeroth order approximation and the corresponding non-standard Lagrangian is given by Eq. (\ref{damped-HOf}) with $L^{f}_{null}=\frac{\dot{x}}{\sqrt{\gamma_{0}x^{2}+c_1}}$. 

On the other hand,  $-\frac{\Phi^{''}}{(\Phi^{'})^3}=\gamma_{0}sin(ax)$, in Eq. (\ref{expanded-eom-damped-oscillator}) generates the non-linear form of Bateman oscillator, i.e., $\ddot{x}+b\dot{x}+\gamma_{0}sin(ax)=0$ in the zeroth order approximation, and therefore the corresponding non-standard Lagrangian is also given by Eq. (\ref{damped-HOf}) with $L^{f}_{null}= \frac{\sqrt{a}\dot{x}}{\sqrt{2}\sqrt{ac_{2}-\gamma_{0}cos(ax)}}$.

\subsection{Duffing Oscillator:}

The above non-standard Lagrangian can easily be extended to a non-standard Lagrangian for the Duffing oscillator by setting $-\frac{\Phi^{''}}{(\Phi^{'})^3}=\gamma_{0}x+\beta x^3$ in Eq. (\ref{expanded-eom-damped-oscillator}) with $\beta$ representing the non-linearity in the restoring force. Then, again,  it follows that $\Phi^{'}=\left[c_{1}+2\gamma_{0}\int^{x}{f(\xi)d\xi}\right]^{-1/2} = \frac{\sqrt{2}}{\sqrt{2\gamma_{0}x^{2}+\beta x^{4}+c_1}}$ and the corresponding non-standard Lagrangian is given by Eq. (\ref{damped-HOf}) with $L^{f}_{null}=\frac{\sqrt{2}\dot{x}}{\sqrt{2\gamma_{0}x^{2}+\beta x^{4}+c_1}}$ in the zeroth order approximation, i.e., $L^{f}_{null} \ll 1$. With the above null Lagrangian, $L^{f}_{null}$, the equation of motion for a damped oscillator, Eq. (\ref{expanded-eom-damped-oscillator}), reduces to the unforced Duffing equation, $\ddot{x}+b\dot{x}+\gamma_{0}x+\beta x^3=0$ in the zeroth order approximation.

Furthermore, the identification of $-\frac{\Phi^{''}}{(\Phi^{'})^3}=\gamma_{0}sin(ax)+\beta x^3$ gives 
$L^{f}_{null}=\frac{\sqrt{2a}\dot{x}}{\sqrt{-4\gamma_{0}cos(ax)+\gamma_{0}\beta x^{4}+ac_1}}$ and the corresponding equation of motion is given by $\ddot{x}+b\dot{x}+\gamma_{0}sin(ax)+\beta x^3=0$ in the zeroth order approximation.

Until now, the scope of applications was restricted by the assumptions $Q=RL_{null}$ and $\lambda=0$. There was no a priori assumption made on the forms of $Q$ and $R$. Hence, it would be curious to explore the effects of further restrictions on both $Q$ and $R$ such that they still satisfy $Q=RL_{null}$, which is presented below.

\subsection{Null Lagrangians and potential functions:}

A convenient assumption would be $Q=L_{null}$ so that $Q=RL_{null}$ implies $R=1$. It is obvious that this assumption will make the non-standard Lagrangians constructed so far depend solely on $L_{null}$. However, as shown in Section 2 and in [23], this further restriction on $Q$ will rather render this construction inexpedient for the generalization to the applications presented above. Nevertheless, it turns out that the application of Corollary 1 does generalize the construction in [23] and furthermore elicits interesting physics.

For instance, with $Q=L_{null}$, $\lambda=0$, $U=U(x,t)$ and $M(L_{null})= m = constant$, the equation of motion in Corollary 1, Eq. (\ref{potential-eom-for-general-lagrangian}), reduces to 
\begin{equation}
 \ddot{x} + \frac{\Phi^{''}}{\Phi^{'}}\dot{x}^{2}+
\frac{mU^{'}}{(\Phi^{'})^2} \left[\frac{L_{null}}{dG/dL_{null}}\right]=0,
\label{eomoscillator-potential}
\end{equation}
where $U^{'}=\partial U/\partial x$. Then, in order to derive an equation of motion for an oscillator, it is necessary to express $\frac{L_{null}}{dG/dL_{null}}$ in Eq. (\ref{eomoscillator-potential}) as follows
\begin{equation}
\frac{dG}{dL_{null}}=-\frac{L_{null}}{1+L^{2}_{null}}.
\label{G-equation-1}
\end{equation}
Upon integration, it follows that $G=c_{1}-ln\sqrt{1+L^{2}_{null}}$ with the constant of integration $c_1$ and the corresponding $F$ is obtained  from Eq. (\ref{FG-relation}) to be $F=tan^{-1}(L_{null})+constant$. Now, inserting Eq. (\ref{G-equation-1}) into Eq. (\ref{eomoscillator-potential}), the equation of motion takes the following form
\begin{equation}
 \ddot{x} + \left(\frac{\Phi^{''}}{\Phi^{'}}-mU^{'}\right)\dot{x}^{2}-
\frac{mU^{'}}{(\Phi^{'})^2} =0,
\label{eomoscillator-potential-1}
\end{equation}
where again $L_{null}=\Phi^{'}\dot{x}$ is used. Next, following the similar steps in Section 4.2, the  exactly the same constraint on the gauge function, $\Phi^{''}+\gamma_{0}f(x)(\Phi^{'})^{3}=0$, has to be satisfied to obtain an equation of motion for an oscillatory system and therefore null-Lagrangians are obtained from $L^{f}_{null}=\dot{x} \Phi^{'}=\dot{x}\left[c_{1}+2\gamma_{0}\int^{x}{f(\xi)d\xi}\right]^{-1/2}$.

However, the appearance of a potential function $U(x,t)$ in the non-standard Lagrangian is the new phenomenon in this construction and $U(x,t)$ must satisfy
\begin{equation}
\frac{\partial U}{\partial x}+\frac{\gamma_{0}}{m}f(x) \left[c_{1}+2\gamma_{0}\int^{x}{f(\xi)d\xi}\right]^{-1}=0.
\label{potential}
\end{equation}
In particular, for a simple harmonic oscillator, the above equation for potential, Eq. (\ref{potential}),
takes the following form

$$ \frac{\partial U}{\partial x}+\frac{\gamma_{0}x}{m(\gamma_{0}x^{2}+c_1)}=0$$ and the potential function is solved to be $U(x) = k-\frac{1}{m}\cdot ln\sqrt{\gamma_{0}x^{2}+c_1}$ with $k$ and $c_1$ being constants of integration. Finally, the corresponding non-standard Lagrangian is given by
\begin{equation}
L_{ns}^{potential}=L_{null}^{f}\cdot tan^{-1}(L_{null}^{f})-ln\sqrt{1+(L_{null}^{f})^2}+U(x),
\label{lns-potential}
\end{equation}
where again $L_{null}=\frac{\dot{x}}{\sqrt{\gamma_{0}x^{2}+c_1}}$. Similarly, the potentials for other oscillators discussed above may be obtained, which is avoided here for the sake of abstaining from repetition.

\section{Discussion}

Though the Null Lagrangians appear irrelevant in the derivation of equation of motion for dynamical systems, the procedure above demonstrates that these innocuous Lagrangians may have applications to the construction of non-standard Lagrangians for important dynamical systems. Moreover, it follows that the appropriate Null Lagrangian for a given dynamical system may be determined from the above construction, which was not possible in the previous construction. For example, the Null Lagrangian for a harmonic oscillator in the above construction is given by $L_{null}=\frac{\dot{x}}{\sqrt{\gamma_{0}x^{2}+c_1}}$, whereas that for the Bateman oscillator turns out to be $L^{f}_{null}= \frac{\sqrt{a}\dot{x}}{\sqrt{2}\sqrt{ac_{2}-\gamma_{0}cos(ax)}}$ with $c_{1}$ and $c_{2}$ being the constants of integration. Therefore, this new generalized procedure enables us to determine a Null Lagrangian for a dynamical system. 

Next, the non-standard Lagrangians (NSLs) constructed in this paper from the null 
Lagrangians are of special forms that must be now compared to the previously 
found NSLs. The form of NSLs originally introduced by Arnold [6] was extensively 
explored in many papers (e.g., [9,10,29-32]).  These NSLs are typically written in 
the following form [30,31]
\begin{equation}
 L_{ns1}(\dot{x},x,t) = \frac{1}{F(x)t) \dot{x} + G(x,t) x + H(x,t)}\ ,
\label{ns1}
\end{equation}
where $F(t,x)$, $G(t,x)$ and $H(t,x)$ are arbitrary, but at least twice 
differentiable, scalar functions of time $t$ and $x$.  The Lagrangian
$L_{ns1}(\dot{x},x,t)$ may also be given in a more general form 
\begin{equation}
L_{ns2}(\dot{x},x,t) = \frac{1}{B(x,t) \dot{x} + C(x,t)}\ ,
\label{ns2}
\end{equation}
where $B(x,t) = F(x,t)$ and $C(x,t) = G(x,t) x + H(x,t)$. An interesting 
result is that the denominator of $L_{ns2}(\dot{x},x,t)$ is the general 
null Lagrangian if, and only if, the following null condition is satisfied 
[23]:
\begin{equation}
\left ( \frac{\partial B(x,t)}{\partial t} \right ) = \frac{\partial [xC(x,t)]}
{\partial x}\ .
\label{nullcond}
\end{equation}
There are many different NSLs of the form given by Eq. (\ref{ns2}) that 
were obtained before for different dynamical systems (e.g., [9,10,29-32]).
However, for most of these NSLs the null condition is not satisfied, which 
means that the relationship between the NSLs and NLs is not valid and,
as a result, most previously obtained NSLs do not have their corresponding 
NLs - see examples presented in Section 2 of this paper as well as examples
given in [23].  These examples clearly show the limitations of the method 
developed in [23]; these limitations are overcome in this paper as the 
presented approach allows finding more general relationships between 
NSLs and NLs, and construct the NSLs for some well-known oscillators
(see Section 4).     

The NSLs derived in this paper by using the newly developed generalized procedure 
described in Section 3 have distinct forms from those obtained before 
[9,10,29-32]. But they are also different from the NSLs obtained by Nucci and Leach 
[8,33,34] and others [35], who constructed them by using the Jacobi Last 
Multiplier method, and those NSLs proposed by El-Nabulsi [7,36-38], who 
applied them to many different physics and astronomy problems.  One 
exception is the NSL originally found by Havas [39] for a harmonic oscillator 
and recently explored in [40], which is a special case of Eq. (\ref{lns-potential}),
when $\gamma_0 = \omega^2$ and $c_1 = 0$, but it is different than the NSL
for a harmonic oscillator derived in this paper and given by Eq. (\ref{HO}); 
see also  Eq. (\ref{HOf}).

Let us point out that the NSLs derived in this paper are significantly different 
in their forms than those presented before in [23].  Moreover, as the results 
of Section 2 show, the previous method based on the relationship $L_{ns} 
(\dot x, x, t) = 1 / L_{null} (\dot x, x, t)$ allows only finding NSLs whose 
forms are strongly limited.  The generalized method developed in this paper
removes these limitations by postulating more general relationshisps between 
$L_{null} (\dot x, x, t)$ and $ L_{ns} (\dot x, x, t)$ (for details, see Section 3 and Section 4) 
and allows deriving a new family of NSLs for some of the best known and 
most commonly used linear and nonlinear oscillators in dynamical systems.

\section{Conclusions} 

The exploration in this paper is focused on the generalization of the applicability of Null Lagrangians in dynamics. The previous method to construct non-standard Lagrangians from null 
Lagrangians [23] is reviewed and its limitations are identified. A new 
generalized procedure that allows constructing non-standard Lagrangians 
from null Lagrangians surpasses those limitations and its validity 
is demonstrated by deriving the non-standard Lagrangians and the equations
of motion resulting from them for a harmonic oscillator, the linear and nonlinear 
Bateman oscillators, and the Duffing oscillator. In addition, the new generalized method allows the Null Lagrangians for their respective dynamical systems to be determined uniquely.
The derived non-standard 
Lagrangians form a new family of Lagrangians in Classical Mechanics.  The 
obtained results show a new role played by the null Lagrangians and their 
corresponding non-standard Lagrangians in describing linear and nonlinear, 
and dissipative and non-dissipative dynamical systems. 

In summary, a humble effort is made to explore the least investigated territory of differential equations and dynamics, the space of Null-Lagrangians, to discover the novel role of these null Lagrangians in dynamics. We believe that this exploration is important because
these null Lagrangians will shed more light on our investigation into the fundamental
nature of dynamics.





 




 



%



\end{document}